\documentclass[final]{aipproc}
\layoutstyle{6x9}
\def\laeq{\raise.2ex\hbox{$<$}\kern-.75em\lower.9ex\hbox{$\sim$}\,}
\def\gaeq{\raise.2ex\hbox{$>$}\kern-.75em\lower.9ex\hbox{$\sim$}\,}

\begin{document}

\title{Gamma-Ray Pulsar Visibility}

\author{Christo Venter}{
  address={Unit for Space Physics, North-West University, Potchefstroom Campus, Private Bag X6001, Potchefstroom, 2520, South Africa}
}

\author{Ocker C.\ de Jager}{
  address={Unit for Space Physics, North-West University, Potchefstroom Campus, Private Bag X6001, Potchefstroom, 2520, South Africa}
}

\author{Adrian Tiplady}{
  address={Hartebeesthoek Radio Astronomy Observatory, PO Box 443, Krugersdorp, 
1740, South Africa},altaddress={Department of Physics \& Electronics, Rhodes University, PO Box 94, 
Grahamstown 6140, South Africa}
}

\begin{abstract}
PSR~J0437-4715 is a millisecond pulsar (MSP) thought to be ``pair formation starved'' (having limited pair cascades due to magnetic photon absorption). Fortunately the general relativistic (GR) electrodynamical model under consideration applicable to this pulsar have few free parameters. We model PSR~J0437-4715's visibility \citep{Venter05}, using a 3D model which incorporates the variation of the GR E-field over the polar cap (PC), taking different observer and inclination angles into account. Using this pulsar as a case study, one may generalize to conducting a pulsar population visibility study. We lastly comment on the role of the proposed South African SKA (Square Kilometre Array) prototype, KAT (Karoo Array Telescope), for GLAST $\gamma$-ray pulsar identification.
\end{abstract}

\maketitle

\section{Introduction}
Since the discovery of the first pulsar \citep{Hewish68}, much theoretical work has been done 
both in a classical \citep[e.g.][]{Goldreich69,Sturrock71,Arons83,Beskin98} and general relativistic (GR) \citep[e.g.][]{Beskin90,Muslimov92,Gonthier94,Sakai03} framework. In the polar cap (PC) scenario \citep[e.g.][]{Baring04} a number of papers have been written concerning the development and investigation of a self-consistent GR pulsar model \citep[e.g.][]{Muslimov97,Harding98,Harding02}, along with its observational implications \citep[e.g.][]{Venter05,Harding05,Frackowiak05}.
In this model, there exists a spindown luminosity $\dot{E}_{\rm rot,\,break}$ below which no screening of the accelerating E-field will take place due to the production of electron-positron pairs from curvature radiation (CR) $\gamma$-ray photons  \citep{Harding02}. Most millisecond pulsars (MSPs) do not have completely screened E-fields \citep{Harding05} (as `canonical' pulsars do), allowing particles streaming from the PC to be accelerated up to high altitudes \citep{Muslimov04}. This implies high pair production attenuation spectral cutoffs \citep{Harding05}, and therefore photons escaping with relatively large CR energies, which promise favorable conditions for observation depending on the $\gamma$-ray flux.
In this paper, we model PSR~J0437-4715 \citep{Johnston93}, a nearby MSP of known mass \citep{VanStraten01}, from which radio \citep[e.g.][]{Manchester95}, X-ray \citep{Becker93,Zavlin02}, EUV \citep{Edelstein95} and UV \citep{Kargaltsev04} radiation have been detected, in addition to an optical counterpart \citep{Bell93,Danziger93,Bailyn93,VanStraten01}. Examining PSR~J0437-4715 as a case study will provide essential information for evaluating pulsar population visibility.

\begin{figure}[t]
  \includegraphics[height=.32\textheight]{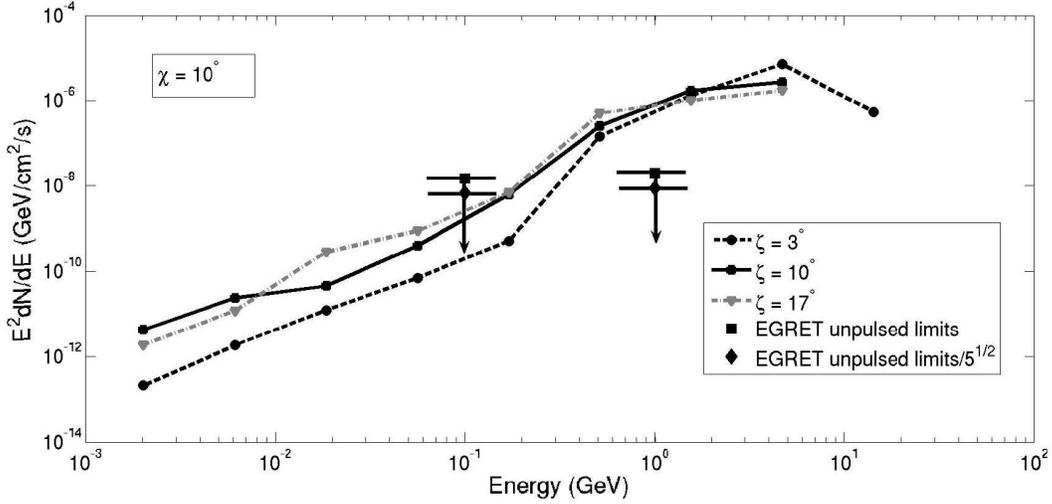}\label{fig1}
  \caption{The $\nu F_{\nu}$-spectrum for different observer angles $\zeta$. Also shown (assuming a $E^{-2}$ photon spectrum) are the EGRET unpulsed upper limits (squares) and pulsed upper limits for a pulse width of 20\% (diamonds).}
\end{figure}

\section{The Model}
The model was described previously \citep{Venter05}. We use the GR B- and E-fields in the frame corotating with the pulsar \citep{Muslimov92,Muslimov97,Harding98,Harding02}. We only consider the dominant CR component of $\gamma$-radiation and use the ``unscreened'' version of the E-field since $\dot{E}_{\rm rot} < \dot{E}_{\rm rot,\,break}$ in this case. We also make allowance for different observer angles $\zeta$ (angle between the rotation axis and the line of sight) and inclination angles $\chi$.
We previously compared PSR~J0437-4715's time-averaged integral flux with the predictions of another pulsar model \citep{Bulik00}, as well as with EGRET upper limits \citep{Fierro95}, and also outlined the role of H.E.S.S.\ in constraining this pulsar's $\gamma$-ray luminosity.

\section{Results}
We find a maximum CR cutoff energy of $1-20$~GeV, depending on the assumed geometry (i.e. $\chi$ and $\zeta$) and equation of state (i.e.\ radius $R$, moment of inertia $I$ and mass $M$). This compares favorably with a value of $\sim$ 10~GeV obtained by \citet{Harding05} and \citet{Frackowiak05}. Our geometry-dependent $\nu F_{\nu}$-spectrum is shown in figure~\ref{fig1} (with $\chi = 10^\circ$). We obtain a harder $\nu F_{\nu}$-spectrum with different spectral maxima than that of \citep{Harding05} and \citep{Frackowiak05}. However, \citet{Harding05} took the scaled magnetic colatitude $\xi = 0.5$, whereas we sampled over a range of $\xi$-values. Furthermore, \citet{Harding05} only calculated a single electron spectrum and normalized it whereas we accelerated primary electrons leaving the stellar surface at a rate of $\sim 10^{31}$ s$^{-1}$ and selected radiation which fell into the range $(\zeta-d\zeta/2,\zeta + d\zeta/2)$.
We converted the EGRET integral upper limits for unpulsed $\gamma$-emission \citep{Fierro95} to differential upper limits assuming a $E^{-2}$ spectrum. It is noted that these upper limits are violated, which indicates necessary refinement of the pulsar model (see also \citep{Venter05} and \citep{Frackowiak05}). 

\begin{figure}[t]
  \includegraphics[height=.32\textheight]{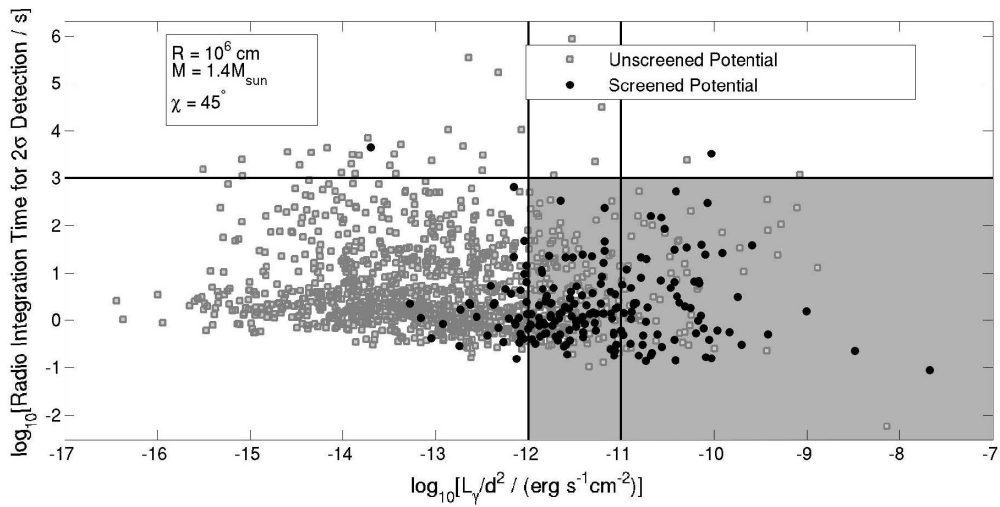}\label{fig2}
  \caption{KAT Radio integration time for 2$\sigma$ detection vs. $L_{\gamma}/d^2$. The region in the lower right corner will be visible to both KAT and GLAST after 1000 seconds of radio integration time.}
\end{figure}

\section{The role of KAT}
We now briefly describe the role a radio telescope will play in the identification of $\gamma$-ray pulsars to be observed by GLAST (\textit{http://www-glast.stanford.edu} and \textit{http://glast.gsfc.nasa.gov}). The proposed Square Kilometre Array (SKA) radio telescope (\textit{http://www.skatelescope.org}) is an international 2~billion dollar project. South Africa is planning the construction of a 1\% prototype (\textit{http://www.ska.ac.za}) named KAT (Karoo Array Telescope) which will be a new generation wide field of view radio telescope (with arcminute resolution) for all sky radio imaging and multiple pulsar surveys. KAT will have  multiple beams (up to 40) allowing it to observe multiple pulsars simultaneously. Since existing radio telescopes cannot monitor all pulsars on a regular basis, most faint $\gamma$-ray pulsars discovered by GLAST would remain unidentified in the absence of contemporary radio parameters. Figure~\ref{fig2} shows the radio integration time vs.\ $L_{\gamma}/d^2$ (see \citep{Harding02} for expressions for $L_{\gamma}$), which is an indication of the expected $\gamma$-flux. We differentiated between the pulsars expected to have screened and unscreened electric potentials. Also shown are 1000 seconds of KAT observation (for $2\sigma$ detection) and the GLAST sensitivity ($10^{-12}-10^{-11}$ erg\,s$^{-1}$\,cm$^{-2}$), boxing the fraction of the pulsar population which will be visible to both KAT and GLAST, and underlining the importance of concurrent radio and $\gamma$-ray observations of pulsars. (Note that KAT integration times of some pulsars may deviate depending on the interstellar medium).

\section{Conclusion}
Recently, \citet{Michel04} severely criticized both the \citet{Goldreich69}-type PC model, as well as outer gap models with gaps of arbitrary sizes and locations. We opt to select a ``clean sample'' of pulsars where the basic B- and E-fields may be fundamentally tested observationally. The ``pair-starved'' (low spindown) pulsar population facilitates investigation of pulsar physics in its simplest form, without the complicating effect of E-field screening evident in younger pulsars. Modeling PSR~J0437-4715 as a case study will provide vital insight for such a pulsar population study which will enable us to draw conclusions regarding galactic pulsar visibility. It is imperative to make model refinements, including the transformation from the corotating to the observer frame, since this should modify some of our results. We lastly note the necessity of contemporary radio parameters when identifying $\gamma$-ray pulsars with GLAST, and we described the role of KAT-type instruments in this regard.

\begin{theacknowledgments}
This publication is based upon work supported by the South African National Research Foundation under
Grant number 2053475.
\end{theacknowledgments}

\bibliographystyle{aipproc}
\bibliography{Master2}

\begin{thebibliography}{30}
\expandafter\ifx\csname natexlab\endcsname\relax\def\natexlab#1{#1}\fi
\providecommand{\enquote}[1]{``#1''}
\expandafter\ifx\csname url\endcsname\relax
  \def\url#1{\texttt{#1}}\fi
\expandafter\ifx\csname urlprefix\endcsname\relax\def\urlprefix{URL }\fi
\providecommand{\eprint}[2][]{\url{#2}}

\bibitem[{Venter} and {de Jager}(2005)]{Venter05}
C.~{Venter}, and O.~C. {de Jager}, \emph{Astrophys. J. Lett.}, \textbf{619},
  L167--L170 (2005), \eprint{astro-ph/0412372}.

\bibitem[{Hewish} et~al.(1968)]{Hewish68}
A.~{Hewish}, S.~J. {Bell}, J.~D. {Pilkington}, P.~F. {Scott}, and R.~A.
  {Collins}, \emph{Nature}, \textbf{217}, 709 (1968).

\bibitem[{Goldreich} and {Julian}(1969)]{Goldreich69}
P.~{Goldreich}, and W.~H. {Julian}, \emph{Astrophys. J.}, \textbf{157}, 869
  (1969).

\bibitem[{Sturrock}(1971)]{Sturrock71}
P.~A. {Sturrock}, \emph{Astrophys. J.}, \textbf{164}, 529 (1971).

\bibitem[{Arons}(1983)]{Arons83}
J.~{Arons}, \emph{Astrophys. J.}, \textbf{266}, 215--241 (1983).

\bibitem[{Beskin} and {Malyshkin}(1998)]{Beskin98}
V.~S. {Beskin}, and L.~M. {Malyshkin}, \emph{MNRAS}, \textbf{298}, 847--853
  (1998), \eprint{astro-ph/9806169}.

\bibitem[{Beskin}(1990)]{Beskin90}
V.~S. {Beskin}, \emph{Soviet Astron. Lett.}, \textbf{16}, 286 (1990).

\bibitem[{Muslimov} and {Tsygan}(1992)]{Muslimov92}
A.~G. {Muslimov}, and A.~I. {Tsygan}, \emph{MNRAS}, \textbf{255}, 61--70
  (1992).

\bibitem[{Gonthier} and {Harding}(1994)]{Gonthier94}
P.~L. {Gonthier}, and A.~K. {Harding}, \emph{Astrophys. J.}, \textbf{425},
  767--775 (1994).

\bibitem[{Sakai} and {Shibata}(2003)]{Sakai03}
N.~{Sakai}, and S.~{Shibata}, \emph{Astrophys. J.}, \textbf{584}, 427--432
  (2003), \eprint{astro-ph/0206502}.

\bibitem[{Baring}(2004)]{Baring04}
M.~G. {Baring}, \emph{Adv. Space Res.}, \textbf{33}, 552--560 (2004),
  \eprint{astro-ph/0308296}.

\bibitem[{Muslimov} and {Harding}(1997)]{Muslimov97}
A.~G. {Muslimov}, and A.~K. {Harding}, \emph{Astrophys. J.}, \textbf{485}, 735
  (1997).

\bibitem[{Harding} and {Muslimov}(1998)]{Harding98}
A.~K. {Harding}, and A.~G. {Muslimov}, \emph{Astrophys. J.}, \textbf{508},
  328--346 (1998), \eprint{astro-ph/9805132}.

\bibitem[{Harding} et~al.(2002)]{Harding02}
A.~K. {Harding}, A.~G. {Muslimov}, and B.~{Zhang}, \emph{Astrophys. J.},
  \textbf{576}, 366--375 (2002), \eprint{astro-ph/0205077}.

\bibitem[{Harding} et~al.(2005)]{Harding05}
A.~K. {Harding}, V.~V. {Usov}, and A.~G. {Muslimov}, \emph{Astrophys. J.},
  \textbf{622}, 531--543 (2005), \eprint{astro-ph/0411805}.

\bibitem[{Fr{\c a}ckowiak} and {Rudak}(2005)]{Frackowiak05}
M.~{Fr{\c a}ckowiak}, and B.~{Rudak}, \emph{Adv. Space Res.}, \textbf{35},
  1152--1157 (2005).

\bibitem[{Muslimov} and {Harding}(2004)]{Muslimov04}
A.~G. {Muslimov}, and A.~K. {Harding}, \emph{Astrophys. J.}, \textbf{617},
  471--479 (2004), \eprint{astro-ph/0408377}.

\bibitem[{Johnston} et~al.(1993)]{Johnston93}
S.~{Johnston}, D.~R. {Lorimer}, P.~A. {Harrison}, M.~{Bailes}, A.~G. {Lyne},
  J.~F. {Bell}, V.~M. {Kaspi}, R.~N. {Manchester}, N.~{D'Amico}, and
  L.~{Nicastro}, \emph{Nature}, \textbf{361}, 613--615 (1993).

\bibitem[{van Straten} et~al.(2001)]{VanStraten01}
W.~{van Straten}, M.~{Bailes}, M.~{Britton}, S.~R. {Kulkarni}, S.~B.
  {Anderson}, R.~N. {Manchester}, and J.~{Sarkissian}, \emph{Nature},
  \textbf{412}, 158--160 (2001), \eprint{astro-ph/0108254}.

\bibitem[{Manchester} and {Johnston}(1995)]{Manchester95}
R.~N. {Manchester}, and S.~{Johnston}, \emph{Astrophys. J. Lett.},
  \textbf{441}, L65--L68 (1995).

\bibitem[{Becker} and {Tr\"{u}mper}(1993)]{Becker93}
W.~{Becker}, and J.~{Tr\"{u}mper}, \emph{Nature}, \textbf{365}, 528 (1993).

\bibitem[{Zavlin} et~al.(2002)]{Zavlin02}
V.~E. {Zavlin}, G.~G. {Pavlov}, D.~{Sanwal}, R.~N. {Manchester}, J.~{Tr{\"
  u}mper}, J.~P. {Halpern}, and W.~{Becker}, \emph{Astrophys. J.},
  \textbf{569}, 894--902 (2002), \eprint{astro-ph/0112544}.

\bibitem[{Edelstein} et~al.(1995)]{Edelstein95}
J.~{Edelstein}, R.~S. {Foster}, and S.~{Bowyer}, \emph{Astrophys. J.},
  \textbf{454}, 442 (1995).

\bibitem[{Kargaltsev} et~al.(2004)]{Kargaltsev04}
O.~{Kargaltsev}, G.~G. {Pavlov}, and R.~W. {Romani}, \emph{Astrophys. J.},
  \textbf{602}, 327--335 (2004), \eprint{astro-ph/0310854}.

\bibitem[{Bell} et~al.(1993)]{Bell93}
J.~F. {Bell}, M.~{Bailes}, and M.~S. {Bessell}, \emph{Nature}, \textbf{364},
  603--605 (1993).

\bibitem[{Danziger} et~al.(1993)]{Danziger93}
I.~J. {Danziger}, D.~{Baade}, and M.~{della Valle}, \emph{Astron. Astrophys.},
  \textbf{276}, 382 (1993).

\bibitem[{Bailyn}(1993)]{Bailyn93}
C.~D. {Bailyn}, \emph{Astrophys. J. Lett.}, \textbf{411}, L83--L85 (1993).

\bibitem[{Bulik} et~al.(2000)]{Bulik00}
T.~{Bulik}, B.~{Rudak}, and J.~{Dyks}, \emph{MNRAS}, \textbf{317}, 97--104
  (2000), \eprint{astro-ph/9912274}.

\bibitem[{Fierro et al.}(1995)]{Fierro95}
J.~M. {Fierro et al.}, \emph{Astrophys. J.}, \textbf{447}, 807 (1995).

\bibitem[{Michel}(2004)]{Michel04}
F.~C. {Michel}, \emph{Adv. Space Res.}, \textbf{33}, 542--551 (2004),
  \eprint{astro-ph/0308347}.

\end{thebibliography}

\end{document}